\documentclass[10pt,journal,compsoc]{IEEEtran}

\ifCLASSOPTIONcompsoc

  \usepackage[nocompress]{cite}
    \usepackage{algorithm}
  \usepackage{multirow}
  \usepackage{graphicx}
  \usepackage{algorithm}
  \usepackage{algorithmic}
  \usepackage{array}
  \usepackage{subfigure}
  \usepackage{amsthm}

  \usepackage[justification=centering]{caption}
  \usepackage{url}
  \usepackage{cite}
  \usepackage{setspace}
  \usepackage{amsmath,amssymb,amsfonts}
  \usepackage{algorithmic}
  \usepackage{graphicx}
  \usepackage{textcomp}
  \usepackage{enumerate}
  \usepackage{amsfonts}
  \usepackage{amsmath} 
  \usepackage{amssymb}
  \usepackage{verbatim}
  \usepackage{float}
  \usepackage{setspace}
  \usepackage{threeparttable}

\else
  \usepackage{cite}
\fi

\ifCLASSINFOpdf

\else

\fi

\begin{document}

\title{ Scientific Workflows in Heterogeneous Edge-Cloud Computing: A Data Placement Strategy Based on Reinforcement learning}

\author{Xin Du,~\IEEEmembership{Student Member,~IEEE,}
        Songtao Tang,~\IEEEmembership{}   
        Zhihui Lu,~\IEEEmembership{Member,~IEEE,}
        Keke Gai,~\IEEEmembership{Senior Member,~IEEE,}
        Jie Wu,~\IEEEmembership{Member,~IEEE}
        and~Patrick C.K. Hung,~\IEEEmembership{Senior Member,~IEEE,}
\IEEEcompsocitemizethanks{\IEEEcompsocthanksitem The preliminary version of this
	work titled “A Novel Data Placement Strategy for Data-Sharing Scientific Workflows in Heterogeneous Edge-Cloud Computing Environments” was
	published in 2020 IEEE International Conference on Web Services (ICWS), Beijing, China, 2020, pp. 498-507~\cite{9284088}. (Corresponding authors: Zhihui Lu.)
\IEEEcompsocthanksitem Xin Du and Songtao Tang are with the School of Computer Science, Fudan University, 200433 Shanghai, China, and also with Engineering Research Center of Cyber Security Auditing and Monitoring, Ministry of Education, 200433 Shanghai, China. E-mail: {xdu20, sttang19}@fudan.edu.cn
\IEEEcompsocthanksitem Zhihui Lu is with the School of Computer Science, Fudan University, 200433 Shanghai, China, and also with Shanghai Blockchain Engineering
Research Center, 200433 Shanghai, China. E-mail: lzh@fudan.edu.cn
\IEEEcompsocthanksitem Jie Wu is with the School of Computer Science, Fudan University, 200433 Shanghai, China, and also with Peng Cheng Laboratory, 518055 Shenzhen, China. E-mail: jwu@fudan.edu.cn
\IEEEcompsocthanksitem Keke Gai is with the School of Cyberspace Security, Beijing Institute of Technology, 100081 Beijing, China. E-mail: gaikeke@bit.edu.cn
\IEEEcompsocthanksitem Patrick C. K. Hung is with Faulty of Business and Information Technology, Ontario Tech University, Canada. E-mail:patrick.hung@uoit.ca.}
\thanks{Manuscript received XXX; revised XXX.}

}

\markboth{Journal of \LaTeX\ Class Files,~Vol.~X, No.~X, August~XXX}%
{Shell \MakeLowercase{\textit{et al.}}: Bare Demo of IEEEtran.cls for Computer Society Journals}

\IEEEtitleabstractindextext{%
\begin{abstract}
The heterogeneous edge-cloud computing paradigm can provide an optimal solution to deploy scientific workflows compared to cloud computing or other traditional distributed computing environments. Owing to the different sizes of scientific datasets and the privacy issue concerning some of these datasets, it is essential to find a data placement strategy that can minimize data transmission time. Some state-of-the-art data placement strategies combine edge computing and cloud computing to distribute scientific datasets. However, the dynamic distribution of newly generated datasets to appropriate datacenters and exiting  the spent datasets are still a challenge during workflows execution. To address this challenge, this study not only constructs a data placement model that includes shared datasets within individual and among multiple workflows across various geographical regions, but also proposes a data placement strategy (DYM-RL-DPS) based on algorithms of two stages. First, during the build-time stage of workflows, we use the discrete particle swarm optimization algorithm with differential evolution to pre-allocate initial datasets to proper datacenters. Then, we reformulate the dynamic datasets distribution problem as a Markov decision process and provide a reinforcement learning-based approach to learn the optimal strategy in the runtime stage of scientific workflows. Through simulating heterogeneous edge-cloud computing environments, we designed comprehensive experiments to demonstrate the superiority of DYM-RL-DPS. The results of our strategy can effectively reduce the data transmission time as compared to other strategies.

\end{abstract}

\begin{IEEEkeywords}
Heterogeneous edge-cloud computing, data-sharing, scientific workflows,  reinforcement learning
\end{IEEEkeywords}}

\maketitle

\IEEEdisplaynontitleabstractindextext

\IEEEpeerreviewmaketitle

\IEEEraisesectionheading{\section{Introduction}\label{sec:introduction}}

\IEEEPARstart{I}{n} recent years, the exponential increase of global cooperation in scientific research and the rapid development of distributed computing technology have resulted in a significant change in scientific applications. They are generally data-intensive and computing-intensive and involve vast interwoven tasks~\cite{kashlev2015typetheoretic}. Because of this, scientific workflow is widely used to represent these complicated applications in several fields such as astronomy, physics, and bioinformatics~\cite{li2015composition}. The datasets of these applications generally have a complex structure and different sizes; hence, the deployment of their scientific workflows has rigid requirements for computational and storage resources. Specifically, the data transmission delay in traditional distributed and cloud computing environments during the execution of scientific workflows is disadvantageous to scientific cooperation. The datasets not only have different characteristics (e.g., private/public datasets), but are also used by or generated for scientific workflows; hence, the data placement should take into account the entire process of scientific workflow in all its different stages (e.g., build-time or runtime). Furthermore, the datasets are often shared among multiple tasks within workflows, including workflows in different geo-distributed organizations. New datasets and tasks generated by scientific workflows need to be deployed dynamically on demand. In addition, private datasets only are allowed to be stored in specific research institutes. To overcome these challenges, an efficient data placement model and strategy must be considered in complex real-life workflows' scenarios.

The emergence of heterogeneous edge-cloud computing provides an optimal paradigm to meet the challenges of a data placement strategy for scientific workflows. Cloud computing, which virtualizes infinite resources with lower maintenance costs, supports large-scale commodity hardware. When deploying scientific workflows, it is high efficiency, flexibility, scalability, cost-efficient but the remote end leads to serious transmission delay and private security problem~\cite{duc2019machine,wu2019qamec,tang2022coordinate}. Edge computing can reduce the data transmission delays and guarantee the copyright and privacy of the scientific datasets. Unfortunately, the limited resources and more expensive cost imply that edge computing resources cannot store all the generated datasets at runtime in a scientific workflow~\cite{zhang2018efficient}. Heterogeneous edge-cloud computing environments, which ensure the resource supply and the security of private datasets~\cite{wu2019mobility,shi2017edge,xiao2020orhrc}, combine the advantages of both edge computing and cloud computing to provide an optimal solution for minimizing the data transmission time. As shown in Fig. 1, the environment covers a lot of datacenters, including cloud datacenters that are distributed geographically and edge micro-datacenters that are near the end-users. Preliminary studies on data placement in heterogeneous edge-cloud computing environments only dealt with static networks and fixed scientific workflows, which are usually produced during the build-time stage in large scientific collaboration projects. However, during the runtime stage of scientific workflows, the mobility of scientific datasets and the dynamics of workflows in complex practical workflows' scenarios should be considered.

In real-world scenarios, a practicable data placement strategy should satisfy the following conditions: First, scientific workflows hold on a large number of collaboration between different research institutes should be distributed and data-intensive. During the execution of workflows, datasets sharing among multiple workflows and tasks is usually employed to improve the efficiency of the whole systems. Furthermore, these scientific datasets and tasks sometimes will be allocated and dispatched between geographically distributed datacenters to facilitate collaborative research. Second, owing to a large number of datasets and complex structures of scientific workflows, combining edge computing and cloud computing ensures high cohesion within a datacenter and low coupling between different datacenters. Third, owing to the special features of confidentiality and copyright protection of some scientific datasets, they cannot be shared through allocation and dispatch. In other words, some geographically distributed research institutes may own their private datasets, which are stored only in their own edge micro-centers. Furthermore, there are significant differences in bandwidth and placement cost between different edge micro-datacenters in different geographic regions. These variations could make a significant influence on the data placement strategy. In summary, reasonable complements of edge computing and cloud computing can be used to optimize the data placement for data-sharing scientific workflows; however, they need an effective mechanism to cooperate with the optimal data placement strategy. 

For the optimal data placement, it is usually formulated as an NP-hard problem with constraints, such as the storage of these datacenters of datasets for different datacenters and so on. Contemporarily, several recent studies have mapped this problem to the NP-hard problem~\cite{xu2019meurep,du2019service,shao2019data,du2019oprc}, but their data placement strategy only considers a workflow execution in a single-region, which not suitable for the actual heterogeneous scientific collaboration environments. Meanwhile, owing to the complex nature and a large number of constraints in heterogeneous edge-cloud environments, the popular research approaches including our original version mainly focus on the built-time stages of scientific workflows, which do not allow new datasets or workflows to be generated/consumed dynamically.

In this study, based on the storage capacity and calculation capacity of different datacenters for data-sharing scientific workflows, we first propose a heuristic algorithm named DE-DPSO-DPA to reduce the data transfer time during the build-time stage. During the execution of multiple workflows in multi-region heterogeneous edge-cloud computing environments, the algorithm not only improves the performance of the whole system, but also considers many factors impacting the data transmission time. To be specific, we consider the number of datacenters, the storage capacity of edge micro-datacenters, the data sharing for multiple workflows in different datacenters, the bandwidth between different datacenters in our data placement model and strategy. Then, we propose a data placement optimization problem during the runtime stages of scientific workflows in the environment. When the newly generated datasets map to their placement and the consumed datasets exit from their placement during the runtime stage, a series of sequential scientific tasks in the scientific workflows involve a large number of dynamic scientific datasets and devices. Some tasks can be accomplished with local datasets whereas others depend on datasets from other micro-datacenters from long-distance regions. When datasets are placed among these datacenters to complete workflow tasks, there is an execution order for them. The most challendge is how to choose the optimal datacenters to place the needed datasets of these workflows for a scientific researcher. Furthermore, we conduct several comprehensive experiments to confirm that our strategy can effectively reduce the data transmission time during the whole process (build-time or runtime stages). 

The main contributions of this study are as follow:
\begin{enumerate}[1)]  
	\item We propose a data-sharing model to simulate the distributed real-world placement scenarios.  There are multiple scientific workflows across different regions, some cloud datacenters, and a large number of edge micro-datacenters. In the heterogeneous environment. We consider the share of datacenters both within tasks and among multiple workflows in our model. In contrast to the previous data placement model, the model also considers the coordination of scientific institute between different regions.	
	\item We first formulate the data placement in the build-time stages of scientific workflows, and propose a data placement algorithm based on combining the advantage of DPSO and DE to distribute the initial datasets. We recode and define the crossover and mutation operator of the DE to better suit our data placement strategy. Through theoretical analysis and simulation experiments, this algorithm is highly efficient to place these datasets in the heterogeneous edge-cloud computing environment.	
	\item Furthermore, we formulate the data placement in the dynamic network as a Markov decision process and propose a reinforcement learning (RL)-based online data placement algorithm named DYM-RL-DPA for optimal placement location. Specifically, the DYM-RL-DPA is a multiple buffer deep deterministic policy gradient algorithm to fully utilize the dynamic characteristics in edge-cloud computing environments. Through simulation experiments and result analysis, it is demonstrated that the proposed dynamic algorithm can achieve a more optimal solution than other state-of-the-art methods.
	\item We propose DYM-RL-DPS, which is a dynamic data placement strategy based on DE-DPSO-DPA and DYM-RL-DPA. To optimize data transmission time in each time slot, we consider plenty of impact factors for scientific workflows in our data placement model. Specially, The impact factors are the number of edge micro-datacenters, the storage capacity of edge micro-datacenters, the data sharing for multiple workflows, and the bandwidth between different datacenters.
\end{enumerate}

The rest of this study is organized as follows. Section 2 demonstrates the motivation example and problem analysis. In section 3, we introduce details of our data placement model and data-sharing workflow notation are defined. Section 4 describes our data placement strategy.  Section 5 describes the analysis of the experimental results. In Section 6, we review the related work. The conclusions  are summarized in Section 7.

\section{Motivating Example and problem analysis}

In this section, we first describe two motivating examples of data-sharing scientific workflows in heterogeneous edge-cloud computing environments. Then, we analyze the motivating examples from different perspectives.

\subsection{Motivating Example}
According to surveys of pulsar searching by Swinburne Astrophysics group (http://astronomy.swinburne.edu.au/), the research process of pulsar searching can be seen as a collaborative, distributed, large-scale scientific workflow by a large amount of geographically distributed organizations such as Parkes Radio Telescope and Goldstone Radio Telescope. There are thousands of radio telescopes generating observation data continuously for researchers from different countries to share. In addition, owing to copyright issues, some institutes have their private datasets, which they must store in their own micro-edge datacenters.

We describe two small-scale motivating examples for data-sharing scientific workflows in astrophysics to illustrate our data placement strategy in the built-time stage and runtime stage, respectively. As shown in Fig. 2 and Fig. 3, in the same region, the simple scenario for the scientific research of pulsar searching in our model have two individual workflows, named workflow 1 and workflow 2. There are 11 tasks $\{t_{1},t_{2},...,t_{11}\}$,  11 datasets $\{d_{1},d_{2},...,d_{11}\}$, and 3 datacenters $\{dc_{1},dc_{2},dc_{3}\}$. Specifically, we set $dc_{1}$ as a cloud datacenter with unlimited storage capacity, and assume $dc_{2}$ and $dc_{3}$ as edge micro-datacenters, which storage capacities are 20 GB and 30 GB, respectively. Due to the bandwidth between the cloud and edge micro-datacenters is much lower than the bandwidth between the edge micro-datacenter, we separately set the bandwidths ${band_{12}, band_{12}, band_{23}}$ across the datacenters as ${10M/s, 20M/s, 150M/s}$ ~\cite{meng2019dedas}.
\begin{table}[!t]
	\centering
	\caption{\\Dataset Size of Data-Sharing Scientific Workflows in Fig. 2}
	\label{tab:performance_comparison}
	\begin{tabular}{p{1.1cm}p{0.2cm}p{0.2cm}p{0.2cm}p{0.2cm}p{0.2cm}p{0.2cm}p{0.2cm}p{0.2cm}p{0.2cm}p{0.2cm}p{0.2cm}}	
		\hline
		Dataset& $d_{1}$& $d_{2}$& $d_{3}$& $d_{4}$& $d_{5}$& $d_{6}$& $d_{7}$& $d_{8}$& $d_{9}$& $d_{10}$& $d_{11}$\\
		Size(GB)& 3.1 & 5.4& 2.1& 1.3& 1.1& 2.3& 1.7& 2.1& 1.5& 0.5& 4.0\\
		\hline
	\end{tabular}
\end{table}
\begin{table}[!t]
	\centering
	\caption{\\Dataset Size of Data-Sharing Scientific Workflows in Fig. 3}
	\label{tab:performance_comparison}
	\begin{tabular}{p{1.1cm}p{0.2cm}p{0.2cm}p{0.2cm}p{0.2cm}p{0.2cm}p{0.2cm}p{0.2cm}p{0.2cm}p{0.2cm}p{0.2cm}p{0.2cm}}	
		\hline
		Dataset& $d_{1}$& $d_{2}$& $d_{3}$& $d_{4}$& $d_{5}$& $d_{6}$& $d_{7}$& $d_{8}$& $d_{9}$& $d_{10}$& $d_{11}$\\
		Size(GB)& 1.5 & 3.0& 1.0& 3.5& 4.0& 2.5& 0.5& 3.0& 2.3& 3.7& 2.5\\
		\hline
	\end{tabular}
\end{table}
 It is worth noting that the datasets in Fig. 2 are not the same as the datasets in Fig. 3. According to the real datasets in astronomy, we list the respective sizes of all the datasets as table 1 and table 2. Because the sizes of workflow tasks are much less than scientific datasets, hence we ignore the transmission time in this study.

In Fig. 2, there are 4 private datasets, which can only be deployed in edge micro-datacenters, are deployed separately in 2 edge micro-datacenters, $dc_{2}$ and $dc_{3}$, and several datasets are shared between different workflows. All datasets are initial datasets, which can be deployed in the built-time stage of scientific workflows. Unlike Fig. 2, Fig. 3 considers the entire process during the execution of scientific workflows in heterogeneous edge-cloud computing environments. As shown in Fig. 3, there are 8 initial datasets and 3 generated datasets; and the initial datasets include 2 shared datasets and 2 private datasets. The private dataset $d_{6}$ only be allowed in edge micro-datacenter $dc_{2}$, and another private dataset $d_{11}$ must be placed in $dc_{3}$. For heterogeneous edge-clould environment as shown in Fig. 3, the number of scientific workflows and datasets covered by micro-edge datacenters changes as some large-scale research projects are considered. On the other hand, in the runtime stage of scientific workflows, with the generation of large amounts of data, some dynamic characteristics need to be considered, such as the increase in the numbers of datasets and workflows, the reduction in resources of micro-edge datacenters, etc. When scientific applications join/exit a large-scale scientific workflow, the data placement strategy has to consider the new placement location of the datasets by cloud or micro-edge datacenters, and which edge micro-datacenters to choose from. 

We not only use the figure 2 and figure 3 to describe the motivating examples in the same region,  but also show that geographically distributed scientific institutes respectively have their own private datasets, workflow systems and data-sharing processes. As shown in Fig. 4, there are 4 workflows, 2 cloud datacenters, and several edge micro-datacenters. Furthermore, there are a lot of datasets and tasks, all of them are scheduled and distributed to these datacenters by data placement strategy. Based on the real scenario, which involves the execution of multiple workflows in multi-region heterogeneous edge-cloud computing environments. In this study, our data-sharing model and strategy are based on multiple cloud datacenters and multiple scientific workflows, we consider some impacts like the bandwidth between datacenters, the number of edge micro-datacenters, and the storage capacity of edge micro-datacenters to find the optimal placement location. 

\subsection{Problem Analysis} 
We analyze the data placement problem of the scientific workflows which are described in section 2.1. 

In the build-time stage of workflows, we first pre-distribute initial datasets and calculate the data transmission time, the amount of data moved, and the number of movements. As shown in Fig. 2(a) and Fig. 2(b), there are two data placement strategies leading to two different results. We first analyze the result of these two strategies shown in Fig. 2. The strategy in Fig. 2(a) requires the datasets to be moved eight times; the amount of data moved was 11.6 GB, the data transfer time was calculated to be 600 s. In Fig. 2(b), we find the strategy has six data movements; the amount of data moved was 8.4 GB, and the data transfer time was approximately 280 s. Furthermore, we show the final placement location of each dataset under these two strategies in Table 3. According to the above analysis, we consider the data placement strategy in Fig. 2(b) is superior to that shown in Fig. 2(a). In the runtime stage of workflows, we dynamically allocate generated datasets to appropriate datacenters as shown in Fig. 3(a) and  Fig. 3(a). We describe the final placement location of each existing dataset at the runtime stage in Table 4 under the two strategies. With the calculation, we obtain the strategy shown in Fig. 3(a) required the datasets to be moved four times, the amount of data moved was 10.6 GB, the data transmission time was calculated to be 350.7 s. On the other hand, in Fig. 4(b), the strategy has three data movements, move 8.3 GB amount of data, and use approximately 335.3 s data transfer time.  In this case, the data placement strategy is shown in Fig. 3(b) is superior to that shown in Fig. 3(a). 
\begin{table}[!t]
	\centering
	\caption{\\The Final Placement Location of Each Dataset in Fig. 2}
	\label{tab:performance_comparison}
	\begin{tabular}{p{1.4cm}p{0.2cm}p{0.2cm}p{0.2cm}p{0.2cm}p{0.2cm}p{0.2cm}p{0.2cm}p{0.2cm}p{0.2cm}p{0.2cm}p{0.2cm}}	
		\hline
		Dataset& $d_{1}$& $d_{2}$& $d_{3}$& $d_{4}$& $d_{5}$& $d_{6}$& $d_{7}$& $d_{8}$& $d_{9}$& $d_{10}$& $d_{11}$\\
		\hline
		$DC_{fig2(a)}$& $dc_{1}$& $dc_{1}$& $dc_{1}$& $dc_{1}$& $dc_{2}$& $dc_{2}$& $dc_{2}$& $dc_{2}$& $dc_{3}$& $dc_{3}$& $dc_{3}$\\
		$DC_{fig2(b)}$& $dc_{1}$& $dc_{1}$& $dc_{2}$& $dc_{1}$& $dc_{2}$& $dc_{3}$& $dc_{2}$& $dc_{2}$& $dc_{3}$& $dc_{3}$& $dc_{3}$\\
		\hline
	\end{tabular}
\end{table}
\begin{table}[!t]
	\centering
	\caption{\\The Final Placement Location of Each Dataset in Fig. 3}
	\label{tab:performance_comparison}
	\begin{tabular}{p{1.4cm}p{0.2cm}p{0.2cm}p{0.2cm}p{0.2cm}p{0.2cm}p{0.2cm}p{0.2cm}p{0.2cm}p{0.2cm}p{0.2cm}p{0.2cm}}	
		\hline
		Dataset& $d_{1}$& $d_{2}$& $d_{3}$& $d_{4}$& $d_{5}$& $d_{6}$& $d_{7}$& $d_{8}$& $d_{9}$& $d_{10}$& $d_{11}$\\
		\hline
		$DC_{fig3(a)}$& $dc_{1}$& $dc_{2}$& $dc_{1}$& $dc_{2}$& $dc_{2}$& $dc_{2}$& $dc_{2}$& $dc_{3}$& $dc_{3}$& $dc_{2}$& $dc_{3}$\\
		$DC_{fig3(b)}$& $dc_{1}$& $dc_{1}$& $dc_{1}$& $dc_{2}$& $dc_{2}$& $dc_{2}$& $dc_{2}$& $dc_{3}$& $dc_{2}$& $dc_{2}$& $dc_{3}$\\
		
		\hline
	\end{tabular}
\end{table}

In Fig. 4, for the sample in Fig 3, we add another geo-distributed organization to simulate the possible real-life scenario more accurately. From Fig. 4, the motivating example involves multiple workflows in multi-region environments. In the model, we consider multiple cloud datacenters and multiple scientific workflows. While proposing a data placement algorithm based on DE-DPSO in the built-time stage, we further propose another data placement algorithm(DYM-RL-DPA) based on reinforcement learning. 

\section{ System model and problem definitions}
In this section, based on the heterogeneous edge-cloud computing environments, we first construct a data placement model and give definitions of data-sharing scientific workflows in detail. Then, we formulate the data placement problem and define the purpose of the data placement strategies.  

\subsection{System Model}
In this study, we build a heterogeneous environment with multiple geographically distributed cloud datacenters and a number of edge micro-datacenters: cloud datacenters generally have unlimited storage resources which only has one in per region , and edge micro-datacenters have limited capacity of storage. We denote the environment as  $DC = \{DC_{c},DC_{e}\}$, and construct the system model. The data placement model consists of $m$ cloud datacenters $DC_{c} = \{ dc_{1}, dc_{2},.., dc_{m}\}$ and $n$ edge micro-datacenters $DC_{e} = \{ dc_{1}, dc_{2},.., dc_{n}\}$. Furthermore, we denote the $i$th datacenter as $dc_{i} = < cap_{i}, type_{i}>$, whose $cap_{i}$ represents its storage capacity and $type_{i}$ represents a flag to distinguish the datacenter is a cloud or an edge micro-datacenter. Specially, when a datacenter is the cloud datacenter, we denote it as $type_{i} = 0$. Others, if $type_{i} = 1$, we consider it  is an edge micro-datacenter. In order to model the dynamic workflow in the runtime stage, we denote scientific applications as  $P=\{p_{1},p_{2},p_{3},...,p_{x}\}$, which can produce the new tasks and generate/consume the datasets. We represent  the bandwidth across different datacenters as $b_{ij}=<band_{ij},type_{i},type_{j}>$, where $band_{ij}$ is the value of the bandwidth and datacenter $i$ is not same as datacenter $j$. Next, we provide the definitions of data-sharing scientific workflows in our model.
 
{\bf {Definition 1.}} Scientific workflow.
In the study, we describe scientific workflows as $W$, and use $\{W_{1}, W_{2},..,W_{l}\}$ to denote the difference of scientific workflows. $l$ is the number of these workflows in the data placement model. We depict every scientific workflow as a directed acyclic graph $W_{k} = (T,R,D)$. Specially, we denote the task set as  $T=\{t_{1},t_{2},...,t_{r}\}$, which has $r$ tasks. we denote the the relationship between tasks as an adjacency matrix $R$. When the task $t_{i}$ has no relationship with another task $t_{j}$, we set $R_{i,j} = 0$. If the task $t_{i}$ precedes another task $t_{j}$,we set $R_{i,j} = 1$. In addition, for these scientific workflows, we represent $DS = \{d_{1},d_{2},...,d_{n}\}$  as all datasets in the model.

{\bf {Definition 2.}} Datasets. Similarly with the state-of-the-art model only requires a scientific workflow, all datasets in our model can fall into two categories: public datasets and private datasets. Due to the confidentiality of scientific datasets or the particularity of some datasets, private datasets cannot be flexibly transferred and allocated from different datacenters. However, for public datasets, different research cooperation institutions can flexibly distribute and share by transferring these datasets from any datacenters in this environment. Hence, we describe a dataset $d_{i}$ as $<dsize_{i}, cn_{i}, dc_{i}, pf_{i},pre_{i},suc_{i},sf_{i}>$. To be specific, we denote the size of a dataset $d_{i}$ as $dsize_{i}$, and use $cn_{i}$ to represent the task set that either generates the dataset $d_{i}$ or needs to be generated from this dataset. When a dataset $d_{i}$ is stored in the datacenter $dc_{i}$, we denote it as $d_{i}.dc_{i}$. We set a flag to indicates whether or not $d_{i}$ is a public dataset, and the flag is denoted as $pf$. If $pf = 0$, the datasets is a public dataset; and if $pf = 1$, it can be deemed a private dataset. When the scientific workflows generate a dataset $d_{i}$, we denote the preceding task set of the dataset as $pre_{i}$, and represent the successive task set which consumes $d_{i}$ as $suc_{i}$. If $pre_{i} = 0$ , we mean $d_{i}$ is an initial dataset. Otherwise, we consider $d_{i}$ is a generated dataset. In addition, in the model, we denote all datasets of data-sharing scientific workflows $W$ as $D_{w}$. We uses the attribute $sf_{i}$ to denote whether or not the dataset $d_{i}$ is shared between different workflows, where $sf_{i}$ = 0 denotes an unshared dataset and $sf_{i}$ = 1 represents a shared dataset.

{\bf {Definition 3.}} Task.  In a scientific workflow $W_{k}$, the task can fall into two categories: input datasets and output datasets. While a task is being executed, the datasets associated with it must have a corresponding location in the model.
We describe the task $t_{i}$ as $<iDS_{i},oDS_{i}, dc_{i}>$. For a task $t_{i}$, we  denotes the input datasets as $iDS_{i}$, and denotes the collection of output datasets as $oDS_{i}$. If the task is scheduled to datacenter $dc_{i}$, we can represent it as $t_{i}.dc_{i}$. Furthermore, we define the task set in workflows $W$ as $T_{w}$. In our model, for the data placement of scientific workflows, we assume the map between the task set and data set is many-to-many, and a task can only be executed if it possesses all the datasets it requires. 

{\bf {Definition 4.}} Data placement map. We define $MS=(W,D,DC,Map)$ as the data placement map, where $Map$ can be denoted as dataset--datacenter mapping. For a scientific workflow $W_{k}$, according to the attibute of datasets, we divide the $Map_{W_{k}}$ into two cases:  private data placement map and public data placement map. We formularize a private data placement map as $Map.pri = \bigcup_{d_{i} \in {d_{i}.pri}}\{d_{i} \to d_{i}.dc \}$, and a public data placement map as $Map.pub = \bigcup_{d_{i} \in {d_{i}.pub}}\{d_{i} \to d_{i}.dc \}$.  We map the datasets in private data placement map directly to their fixed locations (i.e.,edge micro-datacenters). For a public data placement map, we map these datasets by data placement algorithms. 

{\bf {Definition 5.}} data transfer time at build-time. We define the data placement in the model as $S=(W,D,DC,Map,T)$, where $Map$ is represented as dataset--datacenter mapping. The $Map$ = $ <pri,pub>$ denotes global data placement map, which is similar to Definition 4. Furthermore, we further represent $Map.pub$ as $<Map.pub.sh>$ and  $<Map.pub,ush>$, where $Map.pub.sh = \bigcup_{d_{i} \in {d_{i}.pub}}\{d_{i} \to d_{i}.dc | sf_{i} = 1\}$ and $Map.pub.ush = \bigcup_{d_{i} \in {d_{i}.pub}}\{d_{i} \to d_{i}.dc | sf_{i} = 0\}$.
We formulate the private dataset transfer time of workflows $W$ at build-time as:
\begin{equation}
T_{Map.pri}^{BT} = \sum_{t_{r}\in T_{w}}T(Map.pri,t_{r},iDS_{i}.d_{i})
\end{equation}
where $\forall d_{i} \in t_{r}.iDS.pri$, $d_{i}$ is an initial dataset, which $d_{i}.pre_{i}=0$.

For the transfer time of pubic datasets during build-time stages, we consider the data sharing for multiple workflows, thus the transfer time needs to be calculated considering two aspects: shared datasets and unshared datasets. We represent them separately as follows:
\begin{equation}
T_{Map.pub.sh}^{BT} = \sum_{t_{r}\in T_{w}} \sum_{d_{i}\in t_{r}.pub.sh  }^{d_{i}.pre_{i}=0}      T(d_{i},t_{r})
\end{equation}

\begin{equation}
T_{Map.pub.ush}^{BT} = \sum_{t_{r}\in T_{w}} \sum_{d_{j}\in t_{r}.pub.ush  }^{d_{j}.pre_{i}=0}      T(d_{j},t_{r})
\end{equation}

\begin{equation}
T_{Map.pub}^{BT} = T_{Map.pub.sh}^{BT} +  T_{Map.pub.ush}^{BT}
\end{equation}

The data transfer time of scientific workflows $W$ at build-time is:
\begin{equation}
T_{Map}^{BT} = T_{Map.pri}^{BT} +  T_{Map.pub}^{BT}
\end{equation}

{\bf {Definition 6.}} data transfer time at runtime. We describe the calculation method of data transfer time in Definition 5, which is similar to the data transfer time at runtime stage. However, due to the existence of generated datasets, we need to formulate the transfer time at runtime stage as follows:

Firstly, the private dataset transfer time of workflows $W$ at runtime can be denoted as
\begin{equation}
T_{Map.pri}^{RT} = \sum_{t_{r}\in T_{w}}T(Map.pri,t_{r},iDS_{i}.d_{i})
\end{equation}
where $\forall d_{i} \in t_{r}.iDS.pri$, $d_{i}$ is a generated dataset, which $d_{i}.pre_{i} \neq 0$. 

Meanwhile, the public data transfer time of scientific workflows $W$ at runtime is also incurred by shared and unshared datasets:
\begin{equation}
T_{Map.pub.sh}^{RT} = \sum_{t_{r}\in T_{w}} \sum_{d_{i}\in t_{r}.pub.sh  }^{d_{i}.pre_{i}\neq 0}      T(d_{i},t_{r})
\end{equation}

\begin{equation}
T_{Map.pub.ush}^{RT} = \sum_{t_{r}\in T_{wl}} \sum_{d_{j}\in t_{r}.pub.ush  }^{d_{j}.pre_{i}\neq 0}      T(d_{j},t_{r})
\end{equation}

\begin{equation}
T_{Map.pub}^{RT} = T_{Map.pub.sh}^{RT} +  T_{Map.pub.ush}^{RT}
\end{equation}

The data transfer time of scientific workflows $W$ at runtime is:
\begin{equation}
T_{Map}^{RT} = T_{Map.pri}^{RT} +  T_{Map.pub}^{RT}
\end{equation}

{\bf {Definition 7.}} Data transfer time. Finally, we formulate the transfer time of the whole process for data-shared scientific workflows as the sum of time in 
build-time and runtime stages:

\begin{equation}
T_{Map} = T_{Map}^{BT} +  T_{Map.pub}^{RT}
\end{equation}

\subsection{Problem Formulation}

To improve the efficiency of data placement in the model, minimizing the data transmission time, we propose a data placement model that considers data-sharing scientific workflows. The placement model takes into account shared datasets not only within both the individual region and among multiple regions, but also workflows and among multiple workflows. Similar to private datasets, the shared datasets also play an important role in the placement of our model in the whole system. The data-shared placement model is constructed to find good data placement solutions, not from just individual scientific workflow or a geographical region. By placing more shared datasets together, we can better reduce the data placement cost in heterogeneous edge-cloud computing environments. 

Furthermore, instead of placing data in a static network, in this study, the ultimate goal is to find the optimal data placement strategy that minimizes the data transmission time over a sufficiently long time during the execution of the scientific workflow. Generally, the whole process can be divided into some slot times, we denote the data transmission time $T$ at a time slot $t$ as $T_{trans}$.

The $T_{trans}$ can be calculated as follows:
\begin{equation}
T_{trans} = \sum_{i=1}^{\left|DC\right|}\sum_{j\neq i}^{\left|DC\right|}\sum_{k=1}^{\left|D\right|}\dfrac{dsize_{k}}{band_{ij}} \cdot g_{ijk}
\end{equation}
where  $g_{ijk}$ is a parameter, which discern whether a dataset $d_{k}$ is being transferred from different datacenters; If $d_{k}$ is always in a same datacenter, we use $g_{ijk} = 0$ to indicate it. 

The built-time stage of scientific workflows can be seen as a time slot, which only has initial datasets. For the stage, we can formulate the data placement problem in our model as:
\begin{equation}
\begin{cases}
\begin{aligned}
Min \ T_{trans}
\end{aligned}
\\
subject \ to \  \forall i, \sum_{j=1}^{\left|D\right|} d_{j} \cdot l_{ij} \leq cap_{i}
\end{cases}
\end{equation}
where $l_{ij}$ is used as a flag to denote whether the datacenter $dc_{i}$ stores the dataset $d_{j}$. when $l_{ij} = 1$ indicates that the datacenter $dc_{i}$ stores it, and $l_{ij} = 0$ indicates that it does not. we represent the storage capacity of the $i$th datacenter as $cap_{i}$.

However, at the runtime stage, there are an almount of time slots, the core purpose of us is to pursue a minimum average data transmission time while satisfying the storage capacity constraint $cap_{i}$ for each datacenter and the dynamic characteristics $\Theta$. Specially,  $cap_{i,t}$ denotes the storage capacity of the $i$th datacenter at a time slot $t$. The dynamic characteristics $\Theta$ is the placement sequence of datasets in the model at different times, which represents the dynamic change in the location of all data as it is generated and consumed in the workflow. We divide the whole process of the runtime stage for scientific workflows into $V$ time slots and calculate the transfer time $T$ according to the sequence $\Theta$ in any time slot. In runtime stage, the data placement problem can be regarded as:
\begin{equation}
\begin{cases}
\begin{aligned}
Min \ avg \sum_{t=1}^{V}T_{trans}
\end{aligned}
\\

subject \ to \  \forall i, \sum_{j=1}^{\left|D\right|} d_{j} \cdot l_{ij} \leq capacity_{i,t}
\\
subject \ to \  \Theta_{t}
\end{cases}
\end{equation}
where $\Theta_{t}$ denotes the dynamic characteristic at a time slot $t$.
 
For these two stages, we put forward corresponding data placement algorithms respectively. The first is an offline heuristic algorithm where we know the complete information for data-shared scientific workflows in the environments at built-time stage. The second is an online RL-based
algorithm in which we only know the current information for the generated datasets and datacenters. In what follows, we respectively present the offline and online data placement algorithms, and use these two algorithms to synthesize our data placement strategy.
\section{Data Placement Strategy}
In section 3, We have constructed a data-sharing placement model for scientific workflows across multiple geo-distributed regions. Here, we propose a data placement strategy applied to the model. The strategy, named DYM-RL-DSP, provides the method for finding a better data placement map and can minimize data transfer time during the whole process. We describe the data placement strategy with three parts: first, we design a data placement algorithm(DE-DPSO-DPA) to pre-allocate initial datasets to determine the final locations of public datasets during build-time stage. Then, we propose an online data placement algorithm (DYM-RL-DPA) to learn the optimal strategy to dynamically distribute the public datasets at runtime stage of scientific workflows. In the end, we describe our data placement strategy, which is based on the data placement model and data placement algorithms.
\subsection{Build-Time Stage Algorithm}
During the build-time stage, data-sharing scientific workflows in heterogeneous edge-cloud computing environments can be considered as having only constant datasets and workflows. The build-time stage algorithm is used to distribute existing constant datasets to proper datacenters, whose parameters such as storage ability are also fixed. Based on the PSO algorithm and DE algorithm, which proposed by \cite{kennedy1995particle} and \cite{qin2008differential}, our DE-DPSO-DPA is still a heuristic search algorithm. Similar to the PSO, the algorithm can solve discrete problems like data placement by problem coding. Meanwhile, it also combines an efficient global optimization, which same as the DE algorithm.

\begin{algorithm}[!t]
	\caption{DE-DPSO-DPA}
	\label{alg:A}
	\begin{algorithmic}[1]
		\REQUIRE~~\\  T(itermax) , t(current iteration),	n (particles), \\ D (the dimension) , F(scaling factor), $Cr_{g}$, $Cr_{p}$
		
		\ENSURE~~\\ $Res$ (the best optimal solution)
		\STATE Set parameters and Initialize all datasets' placement;
		\STATE 	Set particle dimension $H=|D.pub|$;
		{  \STATE \bf	for}  i : 1 to swarm size n {\bf do}
		{  \STATE \hspace*{0.1in} \bf	for}  d : 1 to $H$ {\bf do}
		\STATE 	\hspace*{0.2in} Initialize $x_{id}^{k}$ randomly;
		{  \STATE \hspace*{0.1in} \bf End for}\\
		\STATE \hspace*{0.1in}	Initialize pbest\\
		{  \STATE  \bf End for}\\
		\STATE 	Initialize gbest\\
		\WHILE{t $<=$ T }
		
		{  \STATE \bf	for}  i = 1 to n {\bf do}
		{\STATE \hspace*{0.1in} Select a, b randomly from particles and a $\neq$ b; }
		{\STATE \hspace*{0.1in} mutation($ x_{i,t-1}, F, x_{a,t-1}, x_{b,t-1}$) by Equation (16)}
		{\STATE \hspace*{0.1in} crossover($x_{pbest,t-1}$,$u_{i,t}$,$Cr_{p}$) by Equation (17)} 
		{\STATE \hspace*{0.1in} crossover($x_{gbest,t-1}$,$u_{i,t}$,$Cr_{g}$) by Equation (17)} 
		{\STATE \hspace*{0.1in} Select $x_{i,t}$ by Equation (18)  }
		{\STATE \bf  end for}	
		{\STATE  t= t+1}		
		\ENDWHILE
		{ \STATE Update the best optimal solution Res} 
		{\STATE  \bf  Output Result}
	\end{algorithmic}
\end{algorithm}
\subsubsection{Problem Encoding} 
According to the coding principle mentioned in~\cite{lin2019time}, to satisfy the well-known obligatory characteristics such as completeness, non-redundancy, and viability, we propose a discrete coding strategy for the data placement problem. Like most meta-heuristic algorithms, our algorithm also generates n-dimensional candidate particles to filter the optimal solution. For data-sharing scientific workflows in our distributed model, a solution of the data placement problem maps a particle. Hence, when the $t$th iteration in our algorithm, the $i$th particle can be formulated as 
\begin{equation}
X_{i,t} = \{x_{i,t}^{1},x_{i,t}^{2},...,x_{i,t}^{d}\}
\end{equation}
where $d$ denotes the dimension of this particle. In the data placement model, we  represents the dimension as the number of datasets. In addition, we denote  the placement location of the $k$th dataset as $x_{i,t}^{k}$ after the $t$th iteration of the algorithm. For a particle in this model, it is worth noting that $Q$ dimensions are represented as the private datasets and  $H$ dimensions are denoted as the datasets shared for multiple workflows.

When placing the existing dataset in the appropriate datacenter, to minimize data transmission time, we propose a data placement algorithm after determining the correspondence between each particle and the candidate solution.
 
\subsubsection{Algorithm  description}
As a classical heuristic algorithm, the update strategy of the PSO algorithm relies on the velocity and position of the particles. However, the disadvantage of the PSO algorithm is easy to fall into local optimization. Furthermore, when the algorithm handles discrete problems like the one in this study, it is often not practicable.

We combine the DE algorithm with PSO algorithm to address these above-mentioned issues. The proposed algorithm DE-DPSO-DPA expands the search capability of the PSO algorithm, and it is also discretized in the data placement scenarios. Algorithm 1 is the pseudocode of our DE-DPSO-DPA algorithm for scientific workflows. Specifically, this algorithm can be divided into two parts: firstly, we initialize all parameters and datasets in the algorithm. In lines 3-9, we preprocess the shared datasets before placing them and update better values by evaluating the current fitness of particles. At the second part(lines 10$\sim$17), at every iteration, we update each particle by adapting the mutation and crossover operators and update the best global solution based on the DE. The update strategy can effectively prevent the algorithm from falling into the local optimal solution. For the $i$th particle at the $t$th iteration, The mutation method can be denoted as follows.
\begin{equation}
u_{i,t} = x_{i,t-1} \oplus F \odot (x_{a,t-1} \ominus x_{b,t-1})
\end{equation}
where the new feasible particle is denoted as $u_{i,t}$, and $F$ represents a scale factor. We generate the new particle by the the mutation, and change the generated particle using a crossover strategy. As shown in Equation (17), we formulate the crossover operator of our algorithm for the individual cognition and social cognition components.
\begin{equation}
\begin{cases}
\vec y = crossover(\vec x_{1}, \vec x_{2},prob)
\\
\vec y[i] =
\begin{cases}

\begin{aligned}
\vec x_{1}[i] \qquad if \quad Random.r < prob
\end{aligned}
\\
\vec x_{2}[i] \qquad \ if \quad Random.r \geq prob
\end{cases}
\end{cases}
\end{equation}
where  $Random.r$ is denoted as a random factor, which between 0 and 1. To control the extent of crossover operations, we set  a parameter as $prob$.
In each algorithm's iteration, we execute twice for the crossover operations to get the optimal solution. Specifically, in the algorithm, we denote $Cr_{p}$ as the crossover parameter to control the distance between the current particle and the local optimal position. Another parameter $Cr_{g}$  can proportionally select indexes in an old particle and replaces the segment between them with the $gbest$ particle segment. $\vec y$, $\vec x_{1}$ and $\vec x_{2}$ also represent particles in different cases. In addition, with the above operations, we obtain a particle $\vec y$ at the $t$th iteration, and denote it as $w_{i,t}$.

The fitness function is usually used to measure the optimality of a particle: the smaller value represents the better the performance. According to the formulation of our data placement problem, for a particle in our algorithm, the better particle should have a smaller data transfer time in the discrete encoding. Hence, we define the fitness value $fit()$ as the data transmission time.

In conclusion, we describe the update method as follows:
\begin{equation}
x_{i,t} =
\begin{cases}
\begin{aligned}
w_{i,t} \qquad if \quad fit(w_{i,t}) < fit(gbest)
\end{aligned} 
\\
x_{i,t-1}\qquad if\quad fit(w_{i,t}) \geq fit(gbest)
\end{cases}
\end{equation}

\subsection{Runtime Stage Algorithm}
When the scientific workflows are at the run-time stage, there are three dynamic information that needs to be considered:  1) the ready tasks, which implies that
parent tasks of these tasks have finished execution and simultaneously all input datasets for these tasks execution have been well prepared,  and are dynamically updated; 2) all datasets generated by tasks are necessarily allocated to appropriate datacenters; 3) owing to the constant changing of amount and location of data in the environment, parameters such as the remaining storage capacity of the edge node are also changing dynamically; 4) as a research project progresses, not only will the number of tasks and workflows in the system change, but the location of the research equipment may also change. The runtime stage algorithm finds a data placement map of the generated datasets, which solves the data placement problem with earlier mentioned dynamic information by learning the reward of current decisions.  
\begin{algorithm}[!t]
	\caption{DYM-RL-DPA}
	\label{alg:A}
	\begin{algorithmic}[1]
		\REQUIRE~~\\  $Q(s,a| \theta ^Q)$ (critic network), $\mu(s| \theta ^\mu)$(actor network),	$\widetilde{Q}$ and 	$\widetilde{\mu}$ (target network) , R[](Replay memory pool), MinT[](the minimum of data transfer time), $\gamma$(Reward discount), $\alpha$ and $\beta$(learning rate), N(time slots)
		
		\ENSURE~~\\ $\Phi(n) $ (learned strategy of optimal scheme)
		\STATE Set parameters and Initialize all networks;
		{  \STATE \bf	for} each epoch i : 1 to M {\bf do}
		\STATE 	\hspace*{0.1in} Initialize $a$ random process $\zeta$ for action exploration;
		\STATE 	\hspace*{0.1in} Initialize initial state $S_{t}^{0}$;
		{  \STATE \hspace*{0.1in} \bf	for} time slot n : 1 to N {\bf do}
		{  \STATE \hspace*{0.3in} \bf	for}  j : 1 to $maxstep$ {\bf do}
		\STATE 	\hspace*{0.5in}  $a_{i,k}^{j}$ = $\mu(s| \theta ^\mu)$ + $ \zeta_{n} $;
		\STATE 	\hspace*{0.5in}  $a_{i,k}^{j}$ $\to$ $s_{i,k}^{j+1}$,  $rw_{i,k}^{j}$ ;
		\STATE 	\hspace*{0.5in}  Update $MinT[]$;
		\STATE 	\hspace*{0.5in}  Store transitions $s_{i,k}^{j},a_{i,k}^{j},rw_{i,k}^{j},s_{i,k}^{j+1}$ in $ R[]$;
		\STATE 	\hspace*{0.5in}  Sample a minibatch from $ R[]$;
		\STATE 	\hspace*{0.5in} $\widetilde{y_{j}}$ = $rw_{i,k}^{j}$ + $\gamma Q^{'}(s_{i,k}^{j+1},\mu ^{'}(s_{i,k}^{j+1}| \theta ^{\mu^{'}})|\theta ^Q)$;
		\STATE 	\hspace*{0.5in} Update critic by Equation (22);
		\STATE 	\hspace*{0.5in} Update the actor policy by Equation (23);
		\STATE 	\hspace*{0.5in} Update the target networks by Equation (24);
		{  \STATE \hspace*{0.3in} \bf End for}\\
	   {  \STATE \hspace*{0.1in} \bf End for}\\
		{  \STATE  \bf End for}\\
	
		{\STATE  \bf  Output Result}
	\end{algorithmic}
\end{algorithm}

\subsubsection{Reinforcement Learning}
As shown in Fig. 5, we describe the interaction of RL agent with the heterogeneous edge-cloud computing environments in this study. At each time slot $t$, the agent observes state of the environment $s_{t}$ and it can be described as $s_{t} = {x(t),h(t)}$ , where $x(t)$ is the edge  micro-datacenter/cloud datacenter that places the datasets and $h(t)$ is the scientific datasets configuration of all datacenters. Furthermore, we take an action $a_{i}$ to obtain the reward $rw_{t}$, and the state $s_{t}$ in the system transmit to  $s_{t+1}$, which is a new state. we describe the process as $<s_{i}, a_{i}, r_{i}, s_{i+1}>$. Furthermore, another action $a_{i+1}$ can also get the corresponding feedback $rw_{t}$, and it is also going to get a  corresponding transformation state $s_{t+1}$. The process can be formulated as a Markov process. The state $s_{t+1}$ is independent of the past steps, and can be determine by  $s_{t}$ and $a_{t}$. Accordingly, we have
\begin{equation}
P(s_{t+1}|s_{t})=P(s_{t+1}|s_{1},s_{2},s_{3},...,s_{t})
\end{equation}
We assume the sequence of this Markov process has $n$ steps. Hence, the sequence can be denoted as
\begin{equation}
<s_{0},a_{0},r_{0},s_{1},a_{1},r_{1},...,s_{n-1},a_{n-1},r_{n-1},s_{n}>
\end{equation}

In addition, we call the probability distribution on the action set as policy $\pi$, which makes control decisions at any state. For RL agents, maximizing the reward is usually the main purpose during each episode. 
By consecutively interacting with the environment, the agent can change the policy during the process to obtain a reward higher than expected by learning. The process of obtaining the higher expected reward is called the training. The data placement strategy can be learned by training.

\subsubsection{RL-based Algorithm}
As shown in Algorithm 2, we propose an RL-based data placement algorithm named DYM-RL-DPA to learn the optimal strategy to find the location of datasets during runtime stage. The objective of our algorithm is to maximize the reward from a memory pool. The reward can
be adjusted by data transmission time as in line 8 in algorithm 2. Furthermore, the output of neutral network is designed as action, which at first explores the environment to accumulate the experience and trains the parameter in the neutral network, finally, we can get the optimized data placement strategy by iteration. Inspired by the Deep Deterministic Policy Gradient (DDPG) algorithm~\cite{9345436}, DYM-RL-DPA has critic network $Q(s,a| \theta ^Q)$ and actor network $\mu(s| \theta ^\mu)$. We denote all parameters in a specific network as $\theta$, and describe the variables in deep Q-network as $\theta^{Q}$. It should be noted that we train the learned strategy in every time slot in each episode, and lines 6$\sim$17 form an iteration in which the agent takes action at current state $s$ to move to the next state $s^{'}$ by receiving an immediate reward. When we sample a minibatch $B$ from the replay memory pool, the reward  $\widetilde{rw}_{i,k}^{j}$ is updated as
\begin{equation}
\widetilde{rw}_{i,k}^{j} = rw_{i,k}^{j} + \gamma Q^{'}(s_{i,k}^{j+1},\mu ^{'}(s_{i,k}^{j+1}| \theta ^{\mu^{'}})|\theta ^Q)
\end{equation}
where $\gamma$ denotes the reward discount, the variables of the target network $\mu^{'}$ are denoted as $\theta^{\mu^{'}}$ in the minibatch. Similarly, in the minibatch, the critic network is described as $Q^{'}$.

We update the critic network by the loss function
\begin{equation}
L=\dfrac{1}{B}\sum_{j}(\widetilde{rw}_{i,k}^{j}- Q(s_{i,k}^j,a_{i,k}^{j}|{\theta^{Q}}))
\end{equation}
and update the actor policy by the following approximation:
\begin{equation}
\bigtriangledown_{\theta}^{\mu} J \approx \dfrac{1}{B} \sum_{j} \bigtriangledown_{a} Q(s,a| \theta ^Q)|_{s={s_{i,k}^{j}},a=\mu(s_{i,k}^{j})} \bigtriangledown_{\theta_{u}}\mu(s|\theta_{u})|s_{i,k}^{j}
\end{equation}
In the end, we update the target networks in both critic
network and actor network with a small constant $\tau$, which is described as follows:
\begin{equation}
\begin{aligned}
\theta^{Q^{'}} \gets \tau\theta^{Q} + (1-\tau)\theta^{Q^{'}}
\\
\theta^{\mu^{'}} \gets \tau\theta^{\mu} + (1-\tau)\theta^{\mu^{'}}
\end{aligned}
\end{equation}

Next, we respectively define the state, the action, and the reward for our data placement problem in this study.

{\bf {State:}} In heterogeneous edge-cloud computing environments, the states need to consider the dynamic characteristic $\theta$ . At time slot $k$, we denote the set of states as $s_{n} = (x(t),h(t))$, here $x(t)$ is the edge  micro-datacenter/cloud datacenter that places the datasets and $h(t)$ is the scientific datasets configuration of all datacenters.  

{\bf {Action:}} In this study, the agent can be considered as DYM-RL-DPA that interconnects with the environment and has learned strategy. At the time slot $t$, the action is a data placement strategy to decide the corresponding datacenters to place the existing datasets. There is a matrix $|DC|\cdot|DS|$ as follows:
$$
\left[
\begin{matrix}
P_{11}      &P_{12}      & \cdots & P_{1|DC|}      \\
P_{21}      & P_{22}      & \cdots &P_{2|DC|}    \\
\vdots & \vdots & \ddots & \vdots \\
P_{|DS|1}     & P_{|DS|2}       & \cdots & P_{|DS||DC|}     \\
\end{matrix}
\right]
$$

{\bf {Reward:}} We propose the data placement strategy to minimize the average of data transmission time in all time slots. According to~\cite{9162056}, at the first iteration, the termination is determined in our design. The reward is set as
\begin{equation}
rw =
\begin{cases}
\begin{aligned}
c_{1} \cdot (T_{avg}(s^j) - T_{avg}(s^{j+1})) \qquad\quad if \quad i = 1
\end{aligned} 
\\

c_{2} + c_{3} \cdot (MinT[k] - T_{avg}(s^{j+1}))\\ \qquad \qquad if \quad i \geq 2, ~ T_{avg}(s^{j+1})< T_{avg} + 1/c_{3}
\\
 c_{4} \cdot (MinT[k] - T_{avg}(s^{j+1})) \qquad \quad if \quad others
\end{cases}
\end{equation}
where $c_{1}$ = 0.1, $c_{2}$ = 1, $c_{3}$ = 0.1 and $c_{4}$ = 0.01 in the reward function.

\begin{algorithm}[!t]
	\caption{DYM-RL-DPS}
	\label{alg:A}
	\begin{algorithmic}[1]
		\REQUIRE~~\\  Datasets in data-sharing scientific workflows $D_{w}$, , Tasks $T_{w}$,  Datacenters $DC$
		\ENSURE~~\\ $PM$ (data placement map), Data transmission time $T_{trans}$	
		\STATE Data transfer time $T_{trans}$ is initialized as 0,  Set queue $RQ$ for ready task and queue $FQ$ for finished tasks;
		\STATE Divide datasets $D_{w}$ into $D_{w}.pub$ and $D_{w}.pri$;
		\STATE Allocate datasets $D_{w}.pri$ to $DC$;
		\STATE Divide datasets $D_{w}.pub$ into $D_{w}.pub.ush$ and  $D_{w}.pub.sh$;
		\STATE During the built-time stage, allocate datasets $D_{w}.pub$ into $DC$ by DE-DPSO-DPA;
		\STATE {\bf While} at runtime  {\bf do}\\	
		\STATE \hspace*{0.2in} Update ready tasks to RQ at runtime;
		\STATE  \hspace*{0.2in} Add all datasets generated by tasks in RQ;
		\STATE \hspace*{0.2in} Divide generated datasets into $D_{w}.pub$ and $D_{w}.pri$;
		\STATE \hspace*{0.2in} Allocate generated private datasets $D_{w}.pri$ to $DC$;
		\STATE \hspace*{0.2in} Divide datasets $D_{w}.pub$ into $D_{w}.pub.ush$ and  $D_{w}.pub.sh$;
		\STATE \hspace*{0.2in} allocate datasets $D_{w}.pub$ into $DC$ by DYM-RL-DPA;
		\STATE  \hspace*{0.2in} Remove all tasks from RQ to FQ;
		\STATE Calculate $T_{trans}$ and assemble $PM$ by Equation(11) ;
		\STATE \bf{Output Result}
	\end{algorithmic}
\end{algorithm}
\subsection{Data Placement Strategy}
We describe the DYM-RL-DPS as shown in algorithm 3, which can use for data-sharing scientific workflows, and consider the entire process during the execution of workflows in heterogeneous edge-cloud environments. To distribute all existing datasets to proper locations, we divide the algorithm 3 into three parts. Firstly, we initialize the data transmission time, ready task queue and finished task queue. With the steps described in 2)$\sim$5) lines, we describe the data placement of initial datasets during built-time stage. Then, data placement of existing  datasets(i.e. initial datasets and generated datasets) are allocated and dispatched during runtime stage (line 6-13). In the end, we combine the former two parts to calculate the data transmission time and assemble the data placement map in line 14.

Specifically, we initialize the current storages of all datacenters, and set the data transmission time as 0. We divide all existing datasets into public datasets and private datasets, and place the private datasets where they are supposed to be. During the runtime stage of scientific workflows, we construct and initialize two queues to cache ready tasks and finished tasks. At the time slot $t$, there may be some new tasks and generated datasets added into $RQ$, meanwhile, some completed tasks and consumed datasets are added into $FQ$. Next, all existing public datasets, which contain unshared and shared datasets, are distributed to appropriate datacenters using the DE-DPSO-DPA at built-time stage and the DYM-RL-DPA at runtime. Finally, we assemble the built-time stage and runtime stage, and calculate the data transmission time until all tasks are completed.

\section{Experimental Results and Analysis}

In this section, we designed comprehensive experiments to evaluate the effectiveness of our proposed data placement strategy. According to the experimental results, we discussed the impact factors in our data placement model. To evaluate the advantages of our proposed strategy, we compared the results with those from other strategies and also considered the different scenarios of these impact factors. We conducted the experiments on a machine with the following specifications: an Intel(R) Core(TM) i7–4790 CPU @ 3.40 GHz, 16 GB of RAM, Windows10(64bit), and IntelliJ IDEA2019.2.4.
	
	
%

\subsection{Experimental Setup}
We used the synthetic workflows from Montage in astronomy released by Bharathi et al~\cite{bharathi2008characterization}, and both the number of datasets and the structures differed in them. In the experiment, the parameters of DE-DPSO-DPA was set as: the initial population size was 100, the maximum number of iterations was 2000, the scaling factor was 0.15, and $Cr_{g}$ and  $Cr_{p}$ were 0.1 and 0.1, respectively. For the DYM-RL-DPA in these experiments, we list the super parameters as follows: both the learning rate for actor network $\alpha$ and the learning rate for critic network $\beta$ were set as 0.001, the reward discount $\gamma$ was 0.99, the constant of the update function for target network $\tau$ was 0.01. In addition, the memory pool of this algorithm was set as 1500, and the batch size was 32.

In the basic experiment, we set two cloud datacenter, and assume only has one cloud datacenter in per region. The cloud datacenter can store  unlimited datasets. In every region,  there are three edge micro-datacenters, which storage capacity is 150 GB. 
We set the bandwidth between different datacenters as follows: 
\begin{equation}
Bandwidth = {
	\left[ \begin{array}{ccccc}
	\sim & 5  & 5  & 5   & 5  \\
	5 & \sim  & 20 &20   &20   \\
	5 & 20 & \sim  &100  &150 \\
	5 & 20 &100 &  \sim &200   \\
	5 & 20 &150 &200  & \sim   \\	
	\end{array} 
	\right ]}
\end{equation}

the bandwidth between two cloud datacenters was set as 5 M/s, the bandwidth between an edge micro-datacenter and a cloud datacenter was 20 M/s, and the bandwidths between different edge micro-datacenters were set as \{100 M/s, 150 M/s, 200 M/s\}. In addition, the experiment covered four workflows and considered the data sharing across scientific workflows.
  
\subsection{Performance Comparison}
Here, we adopt another four data placement strategies to compare with the placement strategies in this study. They are Random-DSP, DE-DSP, DPSO-DSP, GA-DPSO-DSP and DE-DPSO-DSP, respectively. According to their name, we could obtain the algorithm their based on. For example, the DE-DPSO-DSP is based on DE-DPSO algorithm to place the public datasets during the execution of scientific workflows.  All data placement strategies are included the build-time stage and runtime stage. When the strategy is random, during the build time phase, the existing initial public datasets are randomly placed in the datacenter and the initial private datasets are assigned to a specific datacenter. At runtime, the generated public datasets are stored in the data center where the task that generated them resides. When the strategy is based on the algorithm including DE, DPSO, GA-DPSO and DE-DPSO, which means that the data that exists in the system will follow the algorithm when it is placed, both in the initial phase and in the run phase. To be more persuasive, the data transmission time is measured as the average of 100 repeated experiments.  

\subsubsection{Impact of the different bandwidths between edge micro-datacenters}
As shown in Fig.6, for all data placement strategies in these experiments, we set the bandwidth between different edge micro-datacenters as \{0.5, 0.8, 1.5, 3, 5\} times than the bandwidth which is set in the basic experiment. We use these contrasts to compare of the performance different bandwidths across different edge micro-datacenters of different data placement strategies.

\begin{table}[!t]
	\centering
	\fontsize{8}{18}\selectfont
	\caption{Comparison of Time Saving  when Varying
		The	Bandwidth }
	\label{tab:performance_comparison}
	\begin{tabular}{p{2.0cm}c|c|c|c|c}
		
		\hline
		
		\multicolumn{6}{c}{\bf{\qquad\qquad\qquad The Variation of Bandwidth }}\cr\cline{2-6}

		\multirow{1}{*}{\bf{Algorithms}}&
		\multicolumn{1}{c}{\bf{0.5}}&
		\multicolumn{1}{c}{\bf{0.8}}&
		\multicolumn{1}{c}{\bf{1.5}}&
		\multicolumn{1}{c}{\bf{3}}&
		\multicolumn{1}{c}{\bf{5}}\cr\cline{2-6}
		
		\hline
		
		DE-DPS
		&2.57\%
		
		&0.67\%
		
		&0.83\%
		
		&0.64\%
		
		&0.66\%
		
		\cr\hline
		
		GA-DPSO-DPS
		&3.27\%
		
		&1.79\%
		
		&1.13\%
		
		&0.8\%
		
		&0.68\%
		\cr\hline
		DE-DPSO-DPS
		&3.39\%
		
		&2.48\%
		
		&1.14\%
		
		&0.78\%
		
		&0.72\%

		\cr\hline
		{\bf DYM-RL-DPS}&{\bf 11.95\%}&{\bf 7.95\%}&{\bf 13.59\%}&{\bf 9.71\%}&{\bf 19.53\%}\cr\hline
		
	\end{tabular}
\end{table}

Fig.6 shows that the average data transmission time of these six strategies will decrease with the increase of bandwidths between edge micro-datacenters. The Random-DSP is most impacted by the bandwidth, and the differences in results from these strategies based on heuristic algorithms are small. To analyze the percentages of time-saving contributed by all strategies in detail, as listed in Table 5, we compared other strategies with DPSO-DPS, which has the worst performance except Random-DPS.

From Table 5, with the increase of the bandwidth across edge micro-datacenters, the contribution of time-saving offered by our DYM-RL-DPS strategy is always greater than other strategies.

\subsubsection{Impact of the data sharing across workflows}
To validate the importance of our data-sharing placement model, we depict the average data transmission time of different data placement strategies with and without data sharing in Fig. 7. For all these strategies, the average data transfer time using the data-sharing sharing model is smaller than that of the traditional workflow model. Both in the model with and without data sharing, the DYM-RL-DSP can obtain better performance than other strategies. In particular, compared with DE-DPSO-DSP, DYM-RL-DSP reduce the average data transmission time by 6.24\% without data sharing and 11.85\% with data sharing. The result denotes that the proposed DSP method has more advantages in the data-sharing model than other methods. Thereby, joint the data-sharing model and DYM-RL-DSP can make the data placement more high-efficiency.

\subsubsection{Impact of the Different Storage Capacity of Edge micro-Datacenters}
As shown in Fig.8, by varying the storage capacity of edge micro-datacenters from 150G to 350G based on the baseline experiments, we observed and compared the average data transmission time for these strategies. With the increase of the storage capacity of edge micro-datacenters in the environment, the Random-DPS and DPSO-DPS were not very stable, and other strategies kept reduce. The reason is that as the storage capacity of edge micro-datacenters becomes larger, the micro-datacenter can store more datasets. The generated public datasets can be placed in the location where they are generated. When data is no longer being transmitted, there is no data transmission time. 

According to Fig. 8, the performance of our DYM-RL-DSP is always obviously better than other strategies, and the random-DPS is far inferior to others. Specifically, compared with DE-DPSO-DPS, which has the best performance in other strategies, our approach reduces the average data transmission by 9.96\% $\sim$ 14.27\%.

\subsubsection{Impact of the different numbers of edge micro-datacenters}
In Fig.9, we describe the average data transmission time by varying the number of edge micro-datacenters. As per region for our simulation experiments, the number of edge micro-datacenters was set as 3$\sim$5, respectively. 

With the increase of edge micro-datacenter numbers, the figure shows that the average data transmission time was arisen in all strategies. The reason is that as the number of data centers increases, the data placement algorithms of these strategies takes into account more possibilities when placing initial datasets and generated dataset. Furthermore, the performance of our DYM-RL-DSP is always obviously better than other strategies, and the random-DPS is far inferior to others. Specifically, compared with DE-DPSO-DPS, which has the best performance in other strategies, our approach reduces the average data transmission by 8.84\% $\sim$ 11.85\%.


\section{Related Work}
Data placement in distributed systems has been a critical challenge. The strategy for data placement in traditional cloud computing systems has been studied in depth. An efficient data placement strategy can reduce the cost of data storage and network transmission, improving the overall performance of the system. With the development of emerging heterogeneous edge-cloud computing systems, data storage locations have shifted from a centralized cloud to geographically distributed edge-cloud datacenters. A new model is required for new factors in real-world scenarios, such as limited bandwidth and storage capacity of edge micro-datacenters, shared datasets among different edge, and private datasets fixed in specific datacenters. These complex constraints affect the efficiency and solving speed of data placement strategies.

Several previous studies have proposed data placement methods for scientific workflow in traditional distributed computing environments like cluster or grid systems, which can not adapt the characteristics of heterogeneous edge-cloud computing. For example, a data placement strategy with low-level resource sharing will lead to excessive redundant transmission delays between edges. Existing data placement methods mainly have focused on optimizing the simulation models and data transfer time in the cloud. Wang et al.~\cite{wang2014data} proposed a data placement strategy based on k-means clustering for a scientific workflow in cloud environments. This approach focused on the data size and dependency. A data replication mechanism was used for reducing the number of data movements, which did not formalize the data replication cost. Nukarapu et al.~\cite{nukarapu2010data} considered interactions between data placement services and relatively reduced system execution time, designing a classic data-intensive scientific workflow system deployed on a distributed platform. Yuan et al.~\cite{yuan2010data} provided a data placement strategy based on k-means and BEA clustering for a scientific workflow that effectively reduced the number of data movements. However, this method assumed that all edge nodes had the same capacity. Moreover, the number of data movements did not accurately represent the actual data transmission status.

With scientific cloud workflow systems, scientists from different laboratories or regions can collaborate and conduct their research process more flexibly~\cite{li2019overall}. Whereas, in heterogeneous edge-cloud computing systems, datasets locate in geographically distributed data center, resulting in high data transmission cost and long execution time in large-scale scientific workflows. In this paper, a time-driven data placement strategy has been developed by combining the advantages of both edge and cloud computing, which utilizes the computing power of distributed edge micro-datacenters with low data transmission cost.

In summary, existing studies have only focused on individual workflows. The latest data placement models and strategies aim to reduce the data transfer time for individual workflows, ignoring the shared dataset among different workflows. In particular, the discussions on heterogeneous edge-cloud computing environments generally consider only one cloud datacenter and one scientific workflow. Whereas, in practice, cooperation between scientific organizations across different geographical distributions is common. Therefore, data placement models and corresponding strategies should consider multiple scientific workflows among multiple edge cloud datacenters. Under the circumstances, the data sharing will have a significant impact on the performance and the cost of data placement strategies. In addition, while pursuing to improve the performance of the data placement strategy result and reduce the data transmission time, the cost of the solving should also be taken into account. In this study, a novel data placement model is constructed according to the environment mentioned above, and a strategy is proposed to minimize data transfer time for data-sharing scientific workflows.

\section{Conclusion}
In this study, we proposed a dynamic data placement strategy (DYM-RL-DSP), considering the two stages (namely build-time and runtime) of data-sharing scientific workflows in the heterogeneous edge-cloud computing environment. To reduce the data transfer time, we proposed an algorithm named DE-DPSO-DPA, which can pre-allocate initial datasets to appropriate datacenters at the built-time stage. Furthermore, we also proposed a deep reinforcement learning method, DYM-RL-DPA, to optimize dynamic dataset--datacenter mapping during the runtime stage. Based on the two algorithms, DYM-RL-DSP considered large amounts of scientific workflows distributed across different geographic regions, and discussed some impacts such as data sharing between different workflows, the number of edge micro-datacenters, storage capacity of edge micro-datacenters, and bandwidth between datacenters. As compared to other state-of-the-art algorithms, the results of experiments confirm that DYM-RL-DSP can effectively reduce the average data transmission time of scientific workflows in the whole process. For future work, except for scientific workflows, we will investigate other complex data placement problems such as microservice in heterogeneous edge-cloud computing environments, and apply our strategies to solve it. In addition, we will discuss the balance between data transfer time and data placement cost, and consider more variations such as the dataset sizes, number of workflows and the proportion of private datasets, and so on.
\section*{Acknowledgment}
The work of this paper is supported by National Key Research and Development Program of China (2019YFB1405000), National Natural Science Foundation of China under Grant (No. 61873309, No. 61972034,No. 61572137, and No. 61728202), and Shanghai Science and Technology Innovation Action Plan Project under Grant (No.19510710500,No. 18510732000,and No.18510760200).

\bibliographystyle{IEEEtran}
\bibliography{references}

\begin{thebibliography}{10}
\providecommand{\url}[1]{#1}
\csname url@samestyle\endcsname
\providecommand{\newblock}{\relax}
\providecommand{\bibinfo}[2]{#2}
\providecommand{\BIBentrySTDinterwordspacing}{\spaceskip=0pt\relax}
\providecommand{\BIBentryALTinterwordstretchfactor}{4}
\providecommand{\BIBentryALTinterwordspacing}{\spaceskip=\fontdimen2\font plus
\BIBentryALTinterwordstretchfactor\fontdimen3\font minus
  \fontdimen4\font\relax}
\providecommand{\BIBforeignlanguage}[2]{{%
\expandafter\ifx\csname l@#1\endcsname\relax
\typeout{** WARNING: IEEEtran.bst: No hyphenation pattern has been}%
\typeout{** loaded for the language `#1'. Using the pattern for}%
\typeout{** the default language instead.}%
\else
\language=\csname l@#1\endcsname
\fi
#2}}
\providecommand{\BIBdecl}{\relax}
\BIBdecl

\bibitem{9284088}
X.~{Du}, S.~{Tang}, Z.~{Lu}, J.~{Wet}, K.~{Gai}, and P.~C.~K. {Hung}, ``A novel
  data placement strategy for data-sharing scientific workflows in
  heterogeneous edge-cloud computing environments,'' in \emph{2020 IEEE
  International Conference on Web Services (ICWS)}, 2020, pp. 498--507.

\bibitem{kashlev2015typetheoretic}
A.~Kashlev, S.~Lu, and A.~Chebotko, ``Typetheoretic approach to the shimming
  problem in scientific workflows,'' \emph{IEEE Transactions on Services
  Computing}, vol.~8, no.~5, pp. 795--809, 2015.

\bibitem{li2015composition}
H.~Li, K.~C. Chan, M.~Liang, and X.~Luo, ``Composition of resource-service
  chain for cloud manufacturing,'' \emph{IEEE Transactions on industrial
  informatics}, vol.~12, no.~1, pp. 211--219, 2016.

\bibitem{duc2019machine}
T.~L. Duc, R.~G. Leiva, P.~Casari, and P.-O. {\"O}stberg, ``Machine learning
  methods for reliable resource provisioning in edge-cloud computing: A
  survey,'' \emph{ACM Computing Surveys (CSUR)}, vol.~52, no.~5, p.~94, 2019.

\bibitem{wu2019qamec}
Z.~Wu, Z.~Lu, P.~C. Hung, S.-C. Huang, Y.~Tong, and Z.~Wang, ``Qamec: A
  qos-driven iovs application optimizing deployment scheme in multimedia edge
  clouds,'' \emph{Future Generation Computer Systems}, vol.~92, pp. 17--28,
  2019.

\bibitem{tang2022coordinate}
S.~Tang, X.~Du, Z.~Lu, K.~Gai, J.~Wu, P.~C. Hung, and K.-K.~R. Choo,
  ``Coordinate-based efficient indexing mechanism for intelligent iot systems
  in heterogeneous edge computing,'' \emph{Journal of Parallel and Distributed
  Computing}, 2022.

\bibitem{zhang2018efficient}
Q.~Zhang, L.~T. Yang, Z.~Yan, Z.~Chen, and P.~Li, ``An efficient deep learning
  model to predict cloud workload for industry informatics,'' \emph{IEEE
  transactions on industrial informatics}, vol.~14, no.~7, pp. 3170--3178,
  2018.

\bibitem{wu2019mobility}
H.~Wu, S.~Deng, W.~Li, J.~Yin, X.~Li, Z.~Feng, and A.~Y. Zomaya,
  ``Mobility-aware service selection in mobile edge computing systems,'' in
  \emph{2019 IEEE International Conference on Web Services (ICWS)}.\hskip 1em
  plus 0.5em minus 0.4em\relax IEEE, 2019, pp. 201--208.

\bibitem{shi2017edge}
W.~Shi, H.~Sun, J.~Cao, Q.~Zhang, and W.~Liu, ``Edge computing-an emerging
  computing model for the internet of everything era,'' \emph{Journal of
  computer research and development}, vol.~54, no.~5, pp. 907--924, 2017.

\bibitem{xiao2020orhrc}
A.~Xiao, Z.~Lu, X.~Du, J.~Wu, and P.~C. Hung, ``Orhrc: Optimized
  recommendations of heterogeneous resource configurations in cloud-fog
  orchestrated computing environments,'' in \emph{2020 IEEE International
  Conference on Web Services (ICWS)}.\hskip 1em plus 0.5em minus 0.4em\relax
  IEEE, 2020, pp. 404--412.

\bibitem{xu2019meurep}
J.~Xu, X.~Du, W.~Cai, C.~Zhu, and Y.~Chen, ``Meurep: A novel user reputation
  calculation approach in personalized cloud services,'' \emph{PloS one},
  vol.~14, no.~6, p. e0217933, 2019.

\bibitem{du2019service}
W.~Du, T.~Lei, Q.~Het, W.~Liu, Q.~Lei, H.~Zhao, and W.~Wang, ``Service capacity
  enhanced task offloading and resource allocation in multi-server edge
  computing environment,'' in \emph{2019 IEEE International Conference on Web
  Services (ICWS)}.\hskip 1em plus 0.5em minus 0.4em\relax IEEE, 2019, pp.
  83--90.

\bibitem{shao2019data}
Y.~Shao, C.~Li, and H.~Tang, ``A data replica placement strategy for iot
  workflows in collaborative edge and cloud environments,'' \emph{Computer
  Networks}, vol. 148, pp. 46--59, 2019.

\bibitem{du2019oprc}
X.~Du, J.~Xu, W.~Cai, C.~Zhu, and Y.~Chen, ``Oprc: An online personalized
  reputation calculation model in service-oriented computing environments,''
  \emph{IEEE Access}, vol.~7, pp. 87\,760--87\,768, 2019.

\bibitem{kennedy1995particle}
J.~Kennedy and R.~Eberhart, ``Particle swarm optimization (pso),'' in
  \emph{Proc. IEEE International Conference on Neural Networks, Perth,
  Australia}, 1995, pp. 1942--1948.

\bibitem{qin2008differential}
A.~K. Qin, V.~L. Huang, and P.~N. Suganthan, ``Differential evolution algorithm
  with strategy adaptation for global numerical optimization,'' \emph{IEEE
  transactions on Evolutionary Computation}, vol.~13, no.~2, pp. 398--417,
  2008.

\bibitem{lin2019time}
B.~Lin, F.~Zhu, J.~Zhang, J.~Chen, X.~Chen, N.~Xiong, and J.~Lloret, ``A
  time-driven data placement strategy for a scientific workflow combining edge
  computing and cloud computing,'' \emph{IEEE Transactions on Industrial
  Informatics}, 2019.

\bibitem{9345436}
Y.~C. {Liu} and C.~Y. {Huang}, ``Ddpg-based adaptive robust tracking control
  for aerial manipulators with decoupling approach,'' \emph{IEEE Transactions
  on Cybernetics}, pp. 1--14, 2021.

\bibitem{9162056}
Y.~Wang, X.~Wan, X.~Du, X.~Chen, and Z.~Lu, ``A resource allocation strategy
  for edge services based on intelligent prediction,'' in \emph{2021 IEEE 6th
  International Conference on Smart Cloud (SmartCloud)}, 2021, pp. 78--83.

\bibitem{bharathi2008characterization}
S.~Bharathi, A.~Chervenak, E.~Deelman, G.~Mehta, M.-H. Su, and K.~Vahi,
  ``Characterization of scientific workflows,'' in \emph{2008 third workshop on
  workflows in support of large-scale science}.\hskip 1em plus 0.5em minus
  0.4em\relax IEEE, 2008, pp. 1--10.

\bibitem{wang2014data}
M.~Wang, J.~Zhang, F.~Dong, and J.~Luo, ``Data placement and task scheduling
  optimization for data intensive scientific workflow in multiple data centers
  environment,'' in \emph{2014 Second International Conference on Advanced
  Cloud and Big Data}.\hskip 1em plus 0.5em minus 0.4em\relax IEEE, 2014, pp.
  77--84.

\bibitem{nukarapu2010data}
D.~Nukarapu, B.~Tang, L.~Wang, and S.~Lu, ``Data replication in data intensive
  scientific applications with performance guarantee,'' \emph{IEEE Transactions
  on Parallel and Distributed Systems}, vol.~22, no.~8, pp. 1299--1306, 2011.

\bibitem{yuan2010data}
D.~Yuan, Y.~Yang, X.~Liu, and J.~Chen, ``A data placement strategy in
  scientific cloud workflows,'' \emph{Future Generation Computer Systems},
  vol.~26, no.~8, pp. 1200--1214, 2010.

\bibitem{li2019overall}
H.~Li, D.~Yang, W.~Su, J.~L{\"u}, and X.~Yu, ``An overall distribution particle
  swarm optimization mppt algorithm for photovoltaic system under partial
  shading,'' \emph{IEEE Transactions on Industrial Electronics}, vol.~66,
  no.~1, pp. 265--275, 2019.

\end{thebibliography}
%


%




\end{document}